\documentclass[aps,prl,twocolumn,superscriptaddress]{revtex4-1}
\usepackage{units}
\usepackage{graphicx}% Include figure files
\usepackage{dcolumn}% Align table columns on decimal point
\usepackage{bm}
\usepackage{amsmath}% bold math
\begin{document}

% Use the \preprint command to place your local institutional report
% number in the upper righthand corner of the title page in preprint mode.
% Multiple \preprint commands are allowed.
% Use the 'preprintnumbers' class option to override journal defaults
% to display numbers if necessary
%\preprint{}

%Title of paper
\title{Astrophysics of magnetically collimated jets generated from laser-produced
plasmas.}

% repeat the \author .. \affiliation  etc. as needed
% \email, \thanks, \homepage, \altaffiliation all apply to the current
% author. Explanatory text should go in the []'s, actual e-mail
% address or url should go in the {}'s for \email and \homepage.
% Please use the appropriate macro foreach each type of information

% \affiliation command applies to all authors since the last
% \affiliation command. The \affiliation command should follow the
% other information
% \affiliation can be followed by \email, \homepage, \thanks as well.
\author{A. Ciardi}
\affiliation{LERMA, Observatoire de Paris, Universit$\acute{e}$ Pierre et Marie Curie, $\acute{E}$cole Normale Superieure, UMR 8112 CNRS, 5 Place Jules Jannsen, 92195, Meudon, France}

\author{T. Vinci}
\author{J. Fuchs}
\author{B. Albertazzi}
\affiliation{LULI, CNRS, $\acute{E}$cole Polytechnique, Universit$\acute{e}$ Pierre et Marie Curie,  CEA, 91128 Palaiseau, France}

\author{C. Riconda}
\affiliation{LULI, Universit$\acute{e}$ Pierre et Marie Curie, $\acute{E}$cole Polytechnique, CNRS, CEA, 75252, Paris, France}

\author{H. P$\acute{e}$pin}
\affiliation{INRS-EMT, Varennes, Qu$\acute{e}$bec, Canada}

\author{O. Portugall}
\affiliation{Laboratoire National des Champs Magn$\acute{e}$tiques Intenses, Toulouse, France}

%Collaboration name if desired (requires use of superscriptaddress
%option in \documentclass). \noaffiliation is required (may also be
%used with the \author command).
%\collaboration can be followed by \email, \homepage, \thanks as well.
%\collaboration{}
%\noaffiliation

\date{\today}

\begin{abstract}
The generation of astrophysically relevant jets, from magnetically
collimated, laser-produced plasmas, is investigated through three-dimensional,
magneto-hydrodynamic simulations. We show that for laser intensities
$I\sim10^{12}-10^{14}$ W cm$^{-2}$, a magnetic field in excess of $\sim0.1$
MG, can collimate the plasma plume into a prolate cavity bounded by
a shock envelope with a standing conical shock at its tip, which re-collimates
the flow into a super magneto-sonic jet beam. This mechanism is equivalent
to astrophysical models of hydrodynamic inertial collimation, where
an isotropic wind is focused into a jet by a confining circumstellar
torus-like envelope. The results suggest an alternative mechanism
for a large-scale magnetic field to produce jets from wide-angle winds.
\end{abstract}

% insert suggested PACS numbers in braces on next line
% insert suggested keywords - APS authors don't need to do this
%\keywords{}

%\maketitle must follow title, authors, abstract, \pacs, and \keywords
\maketitle

% body of paper here - Use proper section commands
% References should be done using the \cite, \ref, and \label commands

The ejection of mass in the form of bipolar jets is ubiquitous in
astrophysics, and it is widely recognized to be the result of a large-scale,
magnetic field extracting energy from an accreting system\cite{pudritz_magnetic_2012}.
Models of magneto-hydrodynamic (MHD) jet launching \cite{blandford_hydromagnetic_1982}
rely on differential rotation to shear the poloidal magnetic field
component, $\mathbf{B}_{pol}=B_{r}\mathbf{\hat{r}}+B_{z}\mathbf{\hat{z}}$,
and generate a toroidal component, $B_{\theta}$, which is necessary
to power and collimate the flow. The magnetic structure then consists
of helical field lines, with the collimation of the outflow determined
by the component of the Lorentz force perpendicular to $\mathbf{B}_{pol}$,
namely $F_{\perp}=-\frac{B_{\theta}}{\mu_{0}r}\nabla_{\perp}\left(rB_{\theta}\right)+j_{\theta}B_{pol}$.
The first term of $F_{\perp}$ is the essence of self-collimation,
and its role in jet collimation and stability has been studied not
only through multi-dimensional simulations\cite{matsakos_two-component_2009},
but also in experiments using dense, magnetized plasmas\cite{hsu_experimental_2003,lebedev_magnetic_2005,ciardi_episodic_2009}. These experiments can in fact produce flows that are well approximated by the
Euler MHD equations, and whose invariant properties allow meaningful
scaling of laboratory to astrophysical fluid dynamics\cite{ryutov_criteria_2000}. However, collimation solely by the poloidal magnetic field, the term $F_{\perp}\sim j_{\theta}B_{pol}$, still remains to be clarified. For static plasma columns, the confinement was studied in the context of linear theta pinches \cite{Freidberg_1982}. In astrophysics, poloidal collimation can act in the magnetosphere-disc region on scales of a few au (1 astronomical unit  $\sim 1.5\times 10^{13}$ cm), where the collimation of a stellar wind depends
critically on the the magnetic field anchored in the disc\cite{romanova_launching_2009}. On scales of tens of au, it leads to the formation of axially-elongated
cavities\cite{matt_collimation_2003}, and may also serve to re-collimate potentially unstable MHD jets\cite{spruit_collimation_1997}. On even larger scales, it is an essential component of the collimation of outflows embedded in magnetized envelopes\cite{ciardi_outflows_2010}.
\begin{figure}
\includegraphics[width=1\columnwidth]{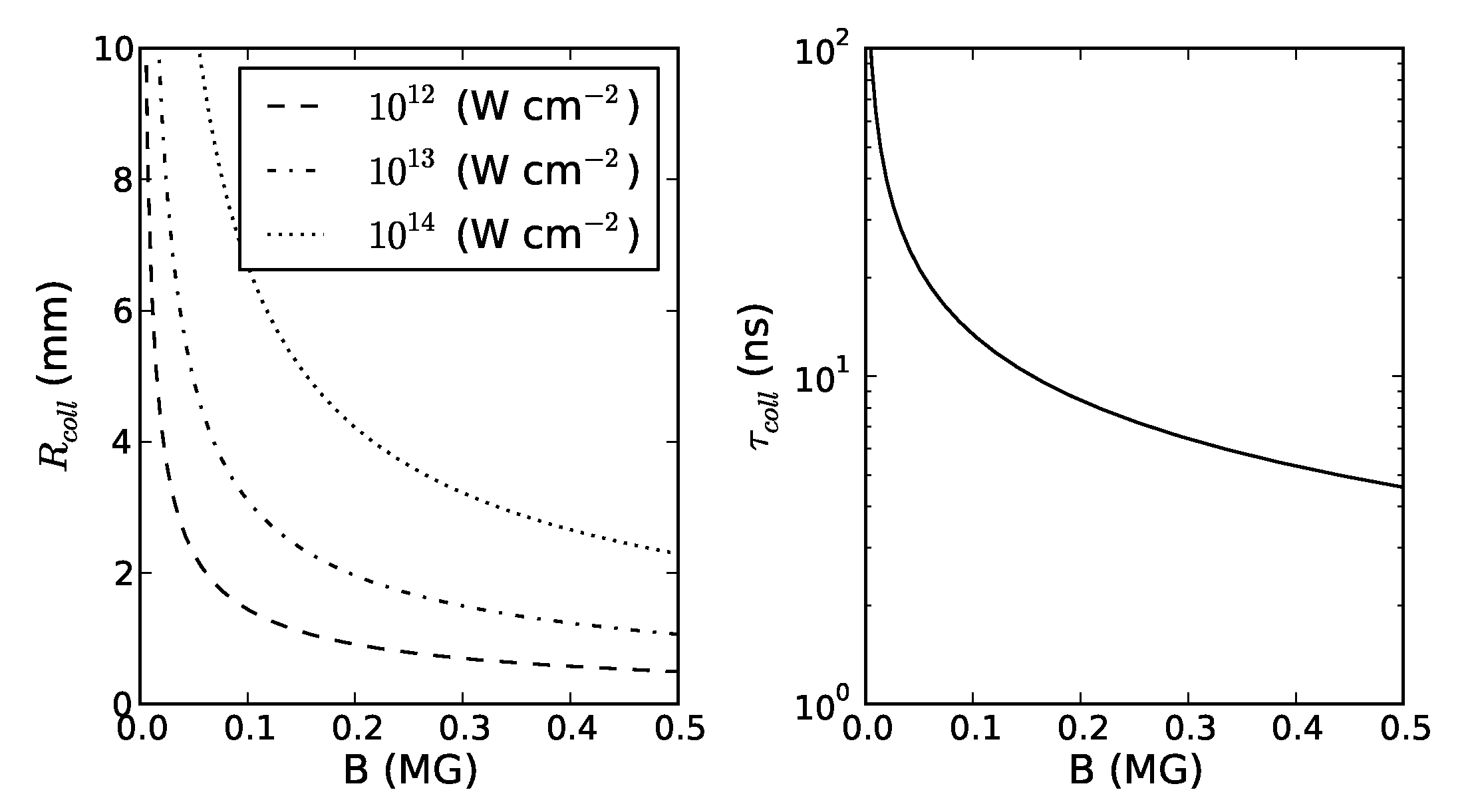}\caption{\label{fig:Initial-collimation-radius}Initial collimation radius
(left) and time-scale (right) calculated
with $f=0.1$. Because of the dependence adopted for $v_{exp}\propto I^{1/3}$, the collimation time-scale is independent of the laser intensity.}
\end{figure}
\begin{figure*}
\includegraphics[width=1\textwidth]{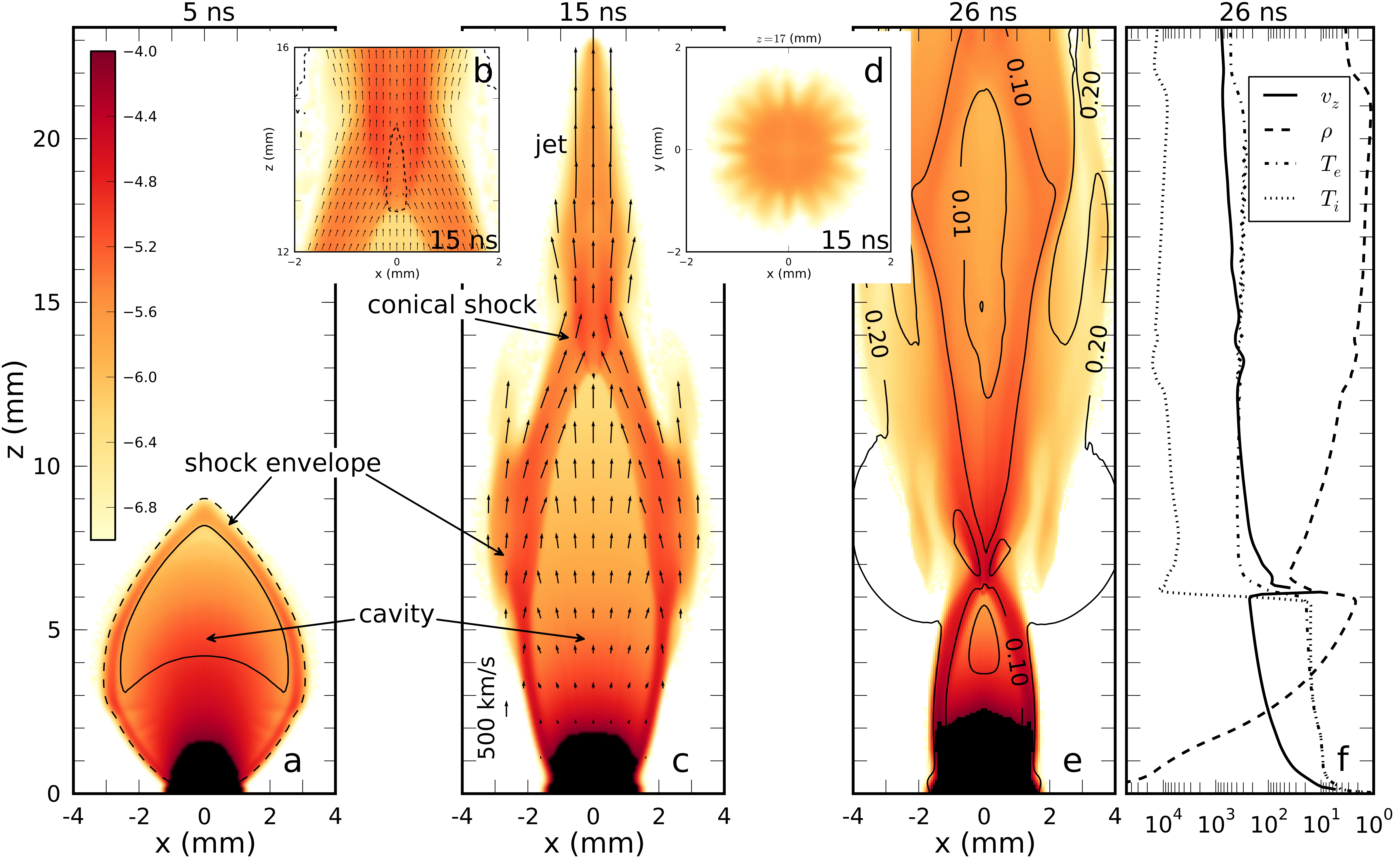}
\caption{\label{fig:Slices} (Colour online) Colour maps correspond to the logarithmic density in g cm$^{-3}$. Panels
(a), (c) and (e) show a cut through the middle of the computational
domain in the $xz$-plane. Contour lines in panel (a) correspond to $M_{ms}=1$ (dashed) and $M_{ms}=10$ (solid).
Velocity vectors are shown in panel (c), while in panel (e) the contours
are for the magnetic field intensity in MG. Panel (b) is a zoom over
the conical shock region depicted in (c), and shows additionally the
region where the flow is sub-magnetosonic, $M_{ms}\leq1$
(dashed line). Panel (d) is a cut perpendicular to the jet at $z=17$ mm. Panel
(f) shows the profiles on axis of density, $\rho\times10^{6}$ (g cm$^{-3}$),
axial velocity, $v_{z}$ (km/s), and ion and electron
temperatures (eV).}
\end{figure*}

In this Letter we establish the astrophysical relevance of coupling
laser-produced plasmas with a strong magnetic field, as a platform
to study jet collimation. Although interest in non-relativistic jet-like flows has instigated a number of experiments using high-power lasers\cite{Remington_Drake_Ryutov_2006}, these remain so far limited
to unmagnetized jets\cite{farley_radiative_1999,loupias_supersonic-jet_2007}. For laser intensities in the range $10^{12}-10^{14}$ W cm$^{-2}$, corresponding to laser energies $E_{L}\sim5-500$ J, with nominal
laser parameters for the pulse duration of $\tau_{L}=1$ ns, focal
spot diameter of $\phi=750$ $\mu$m, and wavelength of $\lambda=1.064$ $\mu$m, we show that under conditions now accessible to current facilities\cite{albertazzi_2012},  a long-duration ($t\gg 10$ ns), strong magnetic field ($>0.1$ MG) can magnetically collimate jet-like flows.  
The basic configuration studied consists of a solid planar target immersed in an externally applied, homogeneous
magnetic field $B_{0}$ parallel to the $z$-axis, and perpendicular
to the target. Using a combination of two- and three-dimensional (3D)
simulations, we provide a theoretical description of the mechanism responsible for generating hydrodynamic jets via a conical shock, from an uncollimated plasma. These results suggest a novel mechanism where wide-angle winds from stars and discs, may be re-collimated into \emph{hydrodynamic} jets by a large-scale, poloidal magnetic field.

For a given applied magnetic field and laser energy, a characteristic collimation radius can be estimated from the equilibrium between ram and magnetic pressures ($\rho v^{2}\sim B_{0}^{2}/8\pi$) as $R_{coll}$(cm)$\sim0.8(E_K$(J)$/B_{0}$(T)$^{2})^{1/3}$, where the bulk kinetic energy is parametrized as a fraction $f$ of the laser energy deposited on target, $E_{K}=fE_{L}$. Numerically we find $f\sim0.3-0.5$, which is consistent with experimental measurement\cite{meyer_experimental_1984}, however, considering only the radial expansion, better
estimates are obtained for$f\sim0.1-0.2$. 
The related collimation time-scale is estimated as $\tau_{coll}\sim R_{coll}/v_{exp}$, where the expansion velocity\cite{tabak_ignition_1994} is $v_{exp}$(cm/s)$=4.6\times10^{7}I^{1/3}\lambda^{2/3}$ ($I$ is the intensity in units of $10^{14}$ W cm$^{-2}$ and $\lambda$ is the laser wavelength in $\mu$m). 
Figure \ref{fig:Initial-collimation-radius} represents $R_{coll}$ and $\tau_{coll}$ as a function of applied magnetic field,  and shows that to magnetically collimate a jet-like flow with a radius of a few millimeters, requires field intensities $\gtrsim0.1$ MG, applied for several tens of nanoseconds.

Numerically, we investigate the interaction of a laser-generated plasma plume
from solid foil targets (C, Al, Cu) with an externally applied, steady-state magnetic field in the range $B_{0}=0-0.4$ MG. Although in this regime a strong (MG) magnetic field can be generated from non-parallel gradients of electronic temperature and density\cite{Stamper_1971}, it remains localized both in time and space\cite{Li_2007,Cecchetti_2009}, and does not affect the plasma dynamics over the time-scales ($\gg \tau_{L} $) and length-scales ($\gg \phi $) of interest to our work. The initial plasma evolution is modelled in axisymmetric, cylindrical geometry with the two-dimensional, three-temperatures, Lagrangian, radiation hydrodynamic code DUED\cite{atzeni_fluid_2005}, coupled with SESAME EOS tables\cite{lyon_1992}. The plasma profiles of density, momentum and temperature (electronic and ionic) are then (at $t=1.2$ ns) linearly mapped onto a 3D Cartesian grid  with a superimposed uniform magnetic field, and used as initial condition for the 3D Eulerian, resistive MHD code GORGON\cite{chittenden_x-ray_2004,ciardi_evolution_2007}. We shall see that 3D calculations are necessary to capture the non-axisymmetric modes of MHD instabilities developing in the flow at late times ($t\gg\tau_{L}$). Simulations were run at different resolutions $(\Delta x=35-65)$ $\mu$m and also with the initial velocity field randomly perturbed ( $\delta v/v_{0}\sim0.05-0.15$). The results are
quantitatively similar, with only small differences in the azimuthal structure of the flow.

\medskip{}

The magnetic collimation of a laser-generated plasma plume may be
characterized by three main phases. These are shown in Fig. \ref{fig:Slices},
for a simulation of an Aluminium target , with $I=1.5\times10^{14}$ W cm$^{-2}$
and $B_{0}=0.2$ MG. The laser propagation is anti-parallel
to the $z$-axis, and the target is at $z=0$ mm. The first
phase (Fig. \ref{fig:Slices}a) corresponds to the initial expansion
of the plasma plume, its deceleration by the radial component of the
Lorentz force $F_{r}=j_{\theta}B_{z}$, and the formation of a shell
of shocked plasma delineating the boundaries of the plume. The time shown ($t=5$ ns)
corresponds approximately to the maximum radial extent ($R_{coll}\sim3-4$ mm)
reached by the thermally-driven expanding plasma. Times are given
from the end of the laser pulse, unless otherwise stated. Because
of the relatively high temperatures, $T_{e}\sim300-500$ eV,
the electrical conductivity is high and dissipation of magnetic flux
via diffusion is small. This is characterized by a relatively high
magnetic Reynolds number, $\textrm{Re}_{M}=1.4\times10^{-20}vL/\eta\sim100$,
where $v\sim10^{7}$ cm/s, $L\sim0.1$ cm
and $\eta\sim1.5\times10^{-16}$ s are the characteristic
velocity, length-scale and resistivity respectively. Therefore the
magnetic field is ``frozen'' in the plasma, and the field lines
are swept laterally by the flow and accumulated in the shock envelope.
In addition, the field lines are bent, generating a radial component of the magnetic
field which produces an additional axial force ($F_{z}=j_{\theta}B_{r}$).
Although this is generally small compared to the thermal pressure
gradients, we shall see later that the curvature of the field lines plays
an important role on the stability of the flow. Velocities in the
plume (few $\times 100$  km/s) are well in excess of
the fast magneto-acoustic speed $c_{ma}$, and the deceleration of
the plasma produces a fast shock; the magneto-acoustic speed, $c_{ma}=\sqrt{c_{A}^{2}+c_{s}^{2}}$,
is a combination of the Alfven speed, $c_{A}=B/\sqrt{4\pi\rho}$,
and the adiabatic sound speed, $c_{s}=\sqrt{\gamma p/\rho}$, with
ratio of specific heats $\gamma=5/3$. 
\begin{figure}[h]
\includegraphics[width=1\columnwidth]{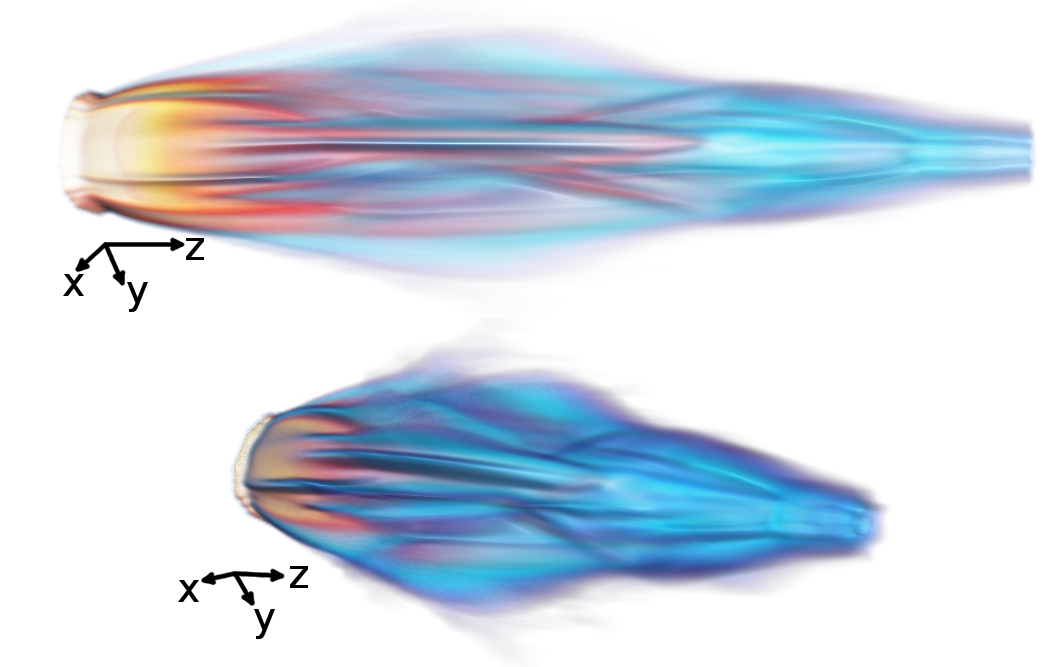}\caption{\label{fig:3D images}(Colour online) Volume rendering of density at 25 ns, showing the structuring of the flow by the RT filamentation instability.}
\end{figure}

%\begin{figure}
%\includegraphics[width=0.6\columnwidth]{images/fig4_column.jpg}
%\caption{\label{fig:varying magnetic field}(Colour online) Line-of-sight integrated
%density, ($\int\rho dy$) in $\unit{g\, cm^{-2}}$ on a logarithmic
%scale, 30 ns after the end of the laser pulse, for an Al target and
%laser intensity $I=1.5\times10^{13}\,\unit{W\, cm^{-2}}$. }
%\end{figure}

 The second phase is the formation of a jet via a standing conical
shock. The propagation in the axial direction is essentially unimpeded
by the magnetic field, and the cavity becomes more elongated in time.
The shock envelope is oblique with respect to the flow
velocity, and compresses the component of the magnetic field tangential
to the shock, while maintaining the tangential velocity continuous.
This axial focusing mechanism is elucidated in Fig. \ref{fig:Slices}c,
where the velocity vectors show the flow being refracted across the
shock, sliding along the walls of the cavity, before finally converging
towards the axis. Furthermore, as demonstrated by the contour lines
of the fast magneto-acoustic Mach number, $M_{ma}=v/c_{ma}$ (Fig.
\ref{fig:Slices}a), the plasma in the shock envelope remains super-fast-magnetosonic.
Its collision on axis can then generate either a conical shock, if
the reflection is regular, or a Mach reflection (Fig. \ref{fig:Slices}b)
consisting of an axisymmetric triple shock structure, with two oblique
(conical) shocks, and a planar shock (Mach disk)\cite{hornung_regular_1986}.
In either case, further acceleration of the flow and its collimation
into a narrow jet occur as the plasma reaching the tip of the cavity
is redirect axially by a conical shock.
The whole flow configuration described so far shares many
important features with astrophysical models of shock focused inertial confinement\cite{balick_shapes_2002}. In those models
the \emph{hydrodynamic} collimation of a (magnetically- or thermally-driven)
wind is the consequence of the inertia of a dense, torus-like circumstellar
envelope, which focuses the flow in the polar direction, forming prolate,
wind-blown cavities, and jets\cite{canto_stellar_1980,frank_hydrodynamical_1996,lee_collimated_2009,Huarte-Espinosa_2012}. Our results show for the first time that an axial magnetic field can in fact mimic the action of a structured, dense envelope, and that the complex physics of jet collimation can be directly accessed in the laboratory. 

Finally, the third phase corresponds to the propagation of the jet, which undergoes one or more expansions
and compressions that may also lead to the further generation of
shocks (interesting similarities exist with jets in ultra-fast accelerative flames in obstructed channels\cite{Bychkov_Valiev_Eriksson_2008}). An example of such re-focusing event can be seen in Fig. \ref{fig:Slices}e,
where the contour lines tracing the magnitude of the magnetic field
show a new region of compression at the tip of the jet ($z\sim23$ mm).
Figure \ref{fig:Slices}f, which illustrates the plasma properties
in the jet, shows the profiles on axis, at 26 ns, of the axial velocity,
electron and ion temperatures, and mass density. The shock-heated
jet has relatively low densities, and thermal equilibration between
the ions and electrons is slow, leading to decoupled temperature.
The jet emerging from the conical shock is aligned with the magnetic
field and it is potentially susceptible to the firehose instability,
which may disrupt the flow through long (axial) wavelength, helical-like
distortions (e.g. \cite{benford_stability_1981}). The condition
of growth requires anisotropic pressures, $P_{\parallel}-P_{\perp}>B^{2}/4\pi$,
where the parallel $P_{\parallel}$ and perpendicular $P_{\perp}$
pressures generally include both the thermal pressure, and the ram
pressure due to the bulk motion of the flow ($\rho v^{2}$). For the
highly supersonic, field-aligned flows of interest here, the parallel
pressure is $P_{\parallel}\sim\rho v^{2}$, and the stability condition,
assuming an isotropic thermal pressure, reduces to $M_{A}^{2}-\nicefrac{\beta}{3}>1$.
Although this is only marginally met in the jet's core, the presence
of a dense, strongly magnetized plasma at larger radii, may provide
the apparent stabilization of the flow\cite{benford_stability_1981}. 
\begin{figure}
\includegraphics[width=1\columnwidth]{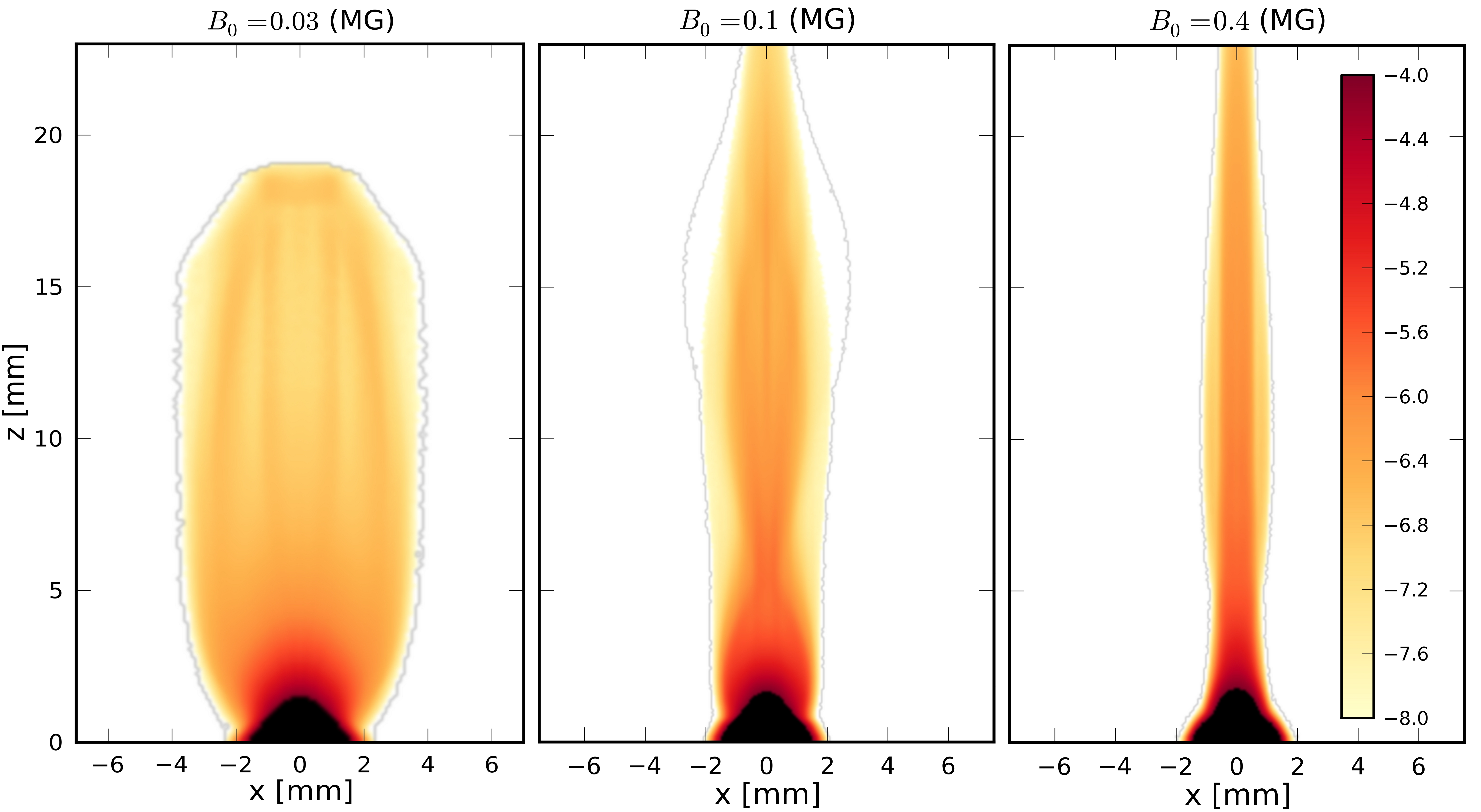}
\caption{\label{fig:varying magnetic field}(Colour online) Line-of-sight integrated
density, ($\int\rho dy$) in g cm$^{-2}$ on a logarithmic
scale, at $t=30$ ns, for an Al target and
laser intensity $I=1.5\times10^{13}$ W cm$^{-2}$. }
\end{figure}
Figure \ref{fig:3D images} shows a three-dimensional view of the
flow at 25 ns. The axial structure consists of alternating regions
where the radius of the flow, $r_{f}(z)$, and curvature of the magnetic
field lines change from convex to concave. In the regions where the
plasma is radially bulging out, a Rayleigh-Taylor type filamentation
instability can develop, with the conditions for its growth being
similar to those of a  theta pinch\cite{kleev_filamentation_1990,haas_stability_1967}.
In particular, the growth rate, $\Gamma$, for large azimuthal mode
numbers $m$, with wavenumber $k_{\theta}=m/r_{f}$, is given by the
classical result $\Gamma\sim\sqrt{gk_{\theta}}$, where $g$ is the
effective gravity at the plasma-vacuum interface, which can be approximated
as $g\sim v_{A}^{2}/R_{c}$, where $R_{c}$ is the radius of curvature\cite{haas_stability_1967}.
In addition, as the flow streams\emph{ }along the curved walls of
the cavity with a velocity $v\gtrsim v_{A}$, it experiences an additional
centrifugal acceleration of the same order of magnitude. Making the
simplifications $R_{c}\sim R_{coll}$ and $v\sim v_{exp}$, which are valid
at early times, shows that the characteristic growth time-scale of
the instability is short, of the order of the collimation time-scale,
\textbf{$\tau_{I}\sim\tau_{coll}/\sqrt{m}$}. These estimates are
consistent with the numerical results, which show azimuthal perturbations,
$m\sim8-16$, growing within a few nanoseconds, and leading to the
rapid filamentation of the outer edges of the cavity first, and of
the jet beam later (see \ref{fig:Slices}d and Fig. \ref{fig:3D images}).
The radially growing perturbations also propagate axially with the
flow (c.f. Fig. \ref{fig:Slices}c), and produce a relatively low
density, broad halo surrounding the central core of the flow.

So far we have discussed the case of a relatively strong magnetic
field, one that is able to generate jets via a conical shock. For
weaker fields in contrast, the flow streamlines tend to become parallel to the
magnetic field lines, and the flow remains instead collimated in a
cylindrical cavity, without jets. The effects that changing the applied
magnetic field has on the collimation and morphology of the flow is
elucidated in Fig. \ref{fig:varying magnetic field}, which
shows for a fixed laser intensity the line-of-sight, integrated mass density. Indeed the flow structure
changes from a cylindrical cavity ($B_{0}\lesssim0.03$ MG), delineated
by a denser shell of plasma, to a prolate cavity with a
jet emerging from a focusing, conical shock. For the largest field
values ($B_{0}\gtrsim0.2$ MG), the focusing conical shock is closer
to the target, and the results is a denser and narrower jet, which
is relatively homogeneous.  We note that by placing a massive target
in the jet propagation path would lead to the formation of a reverse
shock in the jet, in a configuration ideal to study accretion shocks
and magnetized accretion columns in young stars\cite{Orlando_2010}.
Finally, we find that although different target materials lead to
qualitatively similar results, increasing the atomic number of the
target, and thus the radiative losses from the plasma, tends to produce
better collimated jets. This is a well known results from non-magnetized jet
experiments\cite{farley_radiative_1999}. 

The astrophysical relevance of laboratory flows rests on their dynamics being well approximated by ideal MHD\cite{ryutov_criteria_2000}, which  implies the advective transport of momentum, magnetic field, and thermal energy, to dominate over diffusive transport. In this regime the dimensionless Reynolds ($\textrm{Re}$), magnetic Reynolds ($\textrm{Re}_M$), and Peclet ($\textrm{Pe}$) numbers are much larger than unity. The simulations show that the bulk of the flow is well approximated as an ideal magneto-fluid ($\textrm{Re}\sim 10^4 - 10^5$; $\textrm{Re}_{M}\sim100$; $\textrm{Pe}\sim 10 - 20$). Thus we expect astrophysical simulations of related flows, performed under equivalent, scaled initial condition to produce qualitatively similar results. From an astrophysical perspective, the results demonstrate that an axial magnetic field can on its own play the same role of a circumstellar envelope, and lead to flows similar to shock focusing models. Therefore providing an alternative route to explain the presence of jets when massive, collimating envelopes are not consistent with observations\cite{Cabrit_2007}. Moreover, the results suggest a new framework that combines the magnetic collimation of wide-angle
flows with the generation of hydrodynamic jets, which do not suffer
from potentially disruptive instabilities linked to the presence of
a strong $B_{\theta}$. The predicted formation of a standing conical shock is also compelling, as it could possibly explain the presence of stationary emission features observed close to young stellar jet sources\cite{Hartigan_2007, schneider_x-ray_2011}. Although the strength and topology of magnetic fields in those astrophysical jet sources remains a major open question\cite{Hartigan_Frank_Varniére_Blackman_2007}, estimates of its intensity\cite{bonito_x-ray_2011} ($\sim 10$ mG) are consistent with those needed for collimation by a poloidal magnetic field.

% If you have acknowledgments, this puts in the proper section head.
\begin{acknowledgments}
The authors acknowledge the support from the Ile-de-France grant E1127
and from the ANR ``Blanc'' grant SILAMPA.
\end{acknowledgments}

% Create the reference section using BibTeX:
\bibliography{biblio}

\end{document}